# Spectral engineering of integrated photonic filters using mode splitting in silicon nanowire integrated standing-wave resonators

David J. Moss
[a] Centre for Micro-Photonics, Swinburne University of Technology, Hawthorn, 3122, Australia
* Electronic mail: dmoss@swin.edu.au

## ABSTRACT

Mode splitting induced by coherent optical mode interference in coupled resonant cavities is a key phenomenon in photonic resonators that can lead to powerful and versatile filtering functions, in close analogy to electromagnetically-induced-transparency, Autler-Townes splitting, Fano resonances, and dark states. It can not only break the dependence between quality factor, free spectral range, and physical cavity length, but can also lead to group delay response and mode interactions that are useful for enhancing light-material interaction and dispersion engineering in nonlinear optics. In this work, we investigate mode splitting in standing-wave (SW) resonators implemented by cascaded Sagnac loop reflectors (CSLRs) and demonstrate its use for engineering the spectral profile of integrated photonic filters. By changing the reflectivity of the Sagnac loop reflectors (SLRs) and the phase shifts along the connecting waveguides, we tailor mode splitting in the CSLR resonators to achieve a wide range of filter shapes for diverse applications including enhanced light trapping, flat-top filtering, Q factor enhancement, and signal reshaping. We present the theoretical designs and compare the performance of CSLR resonators with three, four, and eight SLRs fabricated in silicon-on-insulator nanowires. We achieve high performance and versatile filter shapes via diverse mode splitting that agree well with theory. The experimental results confirm the effectiveness of our approach towards realizing integrated multi-functional SW filters for flexible spectral engineering.

## 1. INTRODUCTION

Along with the development of advanced micro/nano fabrication technologies, integrated photonic resonators with compact footprints, mass-producibility, high scalability, and versatile filtering properties have been a subject of great interest and become key building blocks for signal modulation, buffering, switching, and processing in optical communication systems [1–3]. Similar to atomic resonance splitting caused by quantum interference between excitation pathways in multi-level atomic systems [4], mode splitting induced by coherent interference between optical modes in coupled resonant cavities is a fundamental phenomenon in integrated photonic resonators, which can lead to many meaningful filtering spectra such as optical analogues of electromagnetically-induced-transparency (EIT), Fano resonances, Autler-Townes splitting (ATS), and dark states [5–7]. Another attractive merit of mode splitting is that it can break the fundamental dependence between quality factor, free spectra range (FSR), and physical cavity length [8–10], which is useful for the implementation of integrated photonic resonators with reduced device footprints and power consumptions. Moreover, the group delay responses and mode interactions generated by mode splitting can also be utilized for enhanced light-material interaction and dispersion engineering in nonlinear optics [11–14].

Generally speaking, photonic resonators can be classified into two categories, namely, travelling-wave (TW) resonators and standing-wave (SW) resonators [15, 16]. The microring resonator (MRR) is a typical example of a TW resonator, while the photonic crystal cavity, distributed feedback cavity, and Fabry-Perot cavity are representatives of SW resonators. To implement integrated photonic resonators with mode splitting, considerable works have been reported [5, 17–22]. Most of these works are based on coupled TW resonators, and in some recent works there are also several device structures constructed by both TW and SW resonators [20, 21]. Considering that the physical cavity length of a TW resonator is almost twice longer as compared with a SW resonator with the same FSR [23], integrated resonators implemented by SW resonators are more attractive in terms of reduced device footprints. In addition, unlike nearly uniform field distribution in TW resonators, there are spacially-dependent field distributions in SW resonators, which are crucial for efficient excitation of laser emission and nonlinear optical effects [16, 24].

In this paper, we investigate mode splitting in integrated photonic resonators implemented by SW resonators. The SW resonators are realized based on multiple cascaded Sagnac loop reflectors (SLRs). The basic idea of photonic resonators formed by cascaded SLRs (we termed as CSLR resonators) was proposed in Ref. [25] and experimental demonstrations of CSLR resonators with two SLRs were reported in Refs. [23, 26]. For CSLR resonators with more than two SLRs, mode splitting occurs due to coherent interference between the FP cavities formed by different SLRs. Here we provide detailed theory of mode splitting in CSLR resonators for spectral engineering and experimental demonstration of CSLR resonators with up to eight SLRs. By changing the reflectivity of the SLRs and the phase shifts along the connecting waveguides, mode splitting in the CSLR resonator can be tailored for diverse applications such as enhanced light trapping, flat-top filtering, Q factor enhancement, and signal reshaping. We present a theoretical analysis for the operation principle, and fabricate the designed CSLR resonators on a silicon-on-insulator (SOI) platform. For the fabricated devices with different numbers of the SLRs, versatile filter shapes that corresponds to diverse mode splitting conditions are experimentally achieved. The experimental results agree well with theory and confirm the effectiveness of the CSLR resonator as a multi-functional SW filter for flexible spectral engineering.

## 2. DEVICE STRUCTURE AND OPERATION PRINCIPLE

Figure 1 illustrates the schematic configuration of the integrated CSLR resonator. It consists of $N$ SLRs (SLR$_1$, SLR$_2$, …, SLR$_N$) formed by a self-coupled nanowire waveguide loop. In the CSLR resonator, each SLR performs as a reflection/transmission element and contributes to the overall transmission spectra from port IN to port OUT in Fig. 1. Therefore, the cascaded SLRs with a periodic loop lattice show similar transmission characteristics to that of photonic crystals [25, 27]. The two adjacent SLRs together with the connecting waveguide form a FP cavity, thereby $N$ cascaded SLRs can also be regarded as $N$-1 cascaded FP cavities (FPC$_1$, FPC$_2$, …, FPC$_{N-1}$), similar to Bragg gratings [16, 28, 29]. To study the CSLR resonator based on the scattering matrix method (SMM) [30–32], we define the waveguide and coupler parameters of the CSLR resonator in Table I. The large dynamic range in engineering the transmittance and reflectivity of individual SLRs via changing $t_i$ or $\kappa_i$ makes the CSLR resonator more flexible for spectral engineering as compared with Bragg gratings. On the other hand, the transmission spectra of the CSLR resonators can also be tailored by changing $\varphi_i$ ($i$ = 1, 2, .., $N$-1), i.e., the phase shifts along the connecting waveguides. The freedom in designing $t_i$ ($i$ =1, 2, .., $N$) and $\varphi_i$ ($i$ = 1, 2, .., $N$-1) is the basis for flexible spectral engineering based on the CSLR resonators, which can lead to versatile applications.

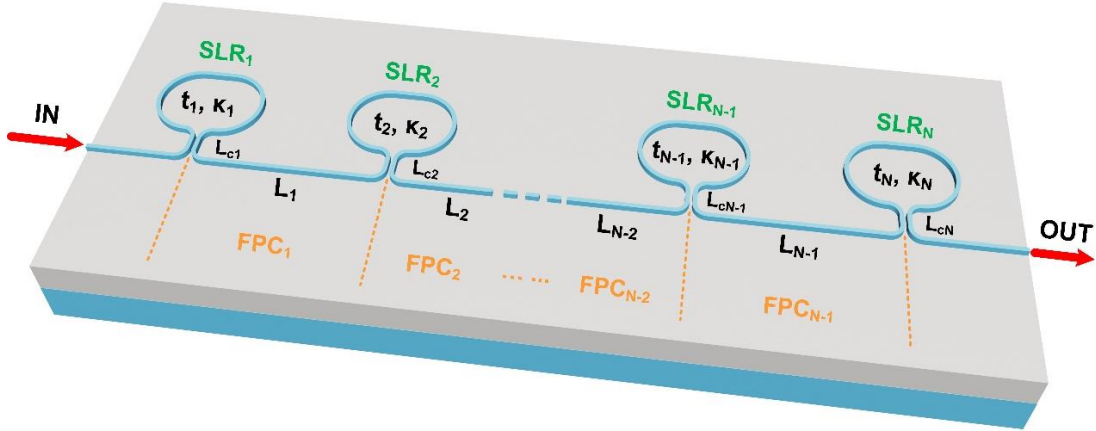

FIG. 1. Schematic configuration of integrated CSLR resonator made up of $N$ cascaded SLRs (SLR$_1$, SLR$_2$, …, SLR$_N$). FPC$_i$ ($i$ = 1, 2, .., $N$-1) are the FP cavities formed by SLR$_i$ and SLR$_{i+1}$, respectively. The definitions of $t_i$ ($i$ = 1, 2, .., $N$), $\kappa_i$ ($i$ = 1, 2, .., $N$), $L_{ci}$ ($i$ = 1, 2, .., $N$), and $L_i$ ($i$ = 1, 2, .., $N$-1) are given in Table I.

CSLR resonators with two SLRs ($N$ = 2) can be regarded as single FP cavities without mode splitting [23, 26]. Here, we start from the CSLR resonators with three SLRs ($N$ = 3). The calculated power transmission spectra and group delay spectra of the CSLR resonators with three SLRs ($N$ = 3) are depicted in Fig. 2. The structural parameters are chosen as follows: $L_{s1} = L_{s2} = L_{s3} = 129.66$ μm, and $L_1 = L_2 = 100$ μm. For single-mode silicon photonic nanowire waveguides with a cross-section of 500 nm × 260 nm, we also assume that the waveguide group index of the transverse electric (TE) mode is $n_g = 4.3350$ and the power propagation loss factor is $\alpha = 55$ m$^{-1}$ (2.4 dB/cm) based on our previously fabricated devices. The same $n_g$ and $\alpha$ are also employed for the calculations of other transmission and group delay spectra in this section. It is clear that different degrees of mode splitting can be achieved by varying $t_2$. As $t_2$ decreases (i.e., the coupling strength increases), the spectral range between the two adjacent resonant peaks decreases until the split peaks finally merge into one. By further decreasing $t_2$, the Q factor, extinction ratio, and group delay of the combined single resonance increases, together with an increase in the insertion loss. In particular, when $t_2 = 0.77$, a band-pass Butterworth filter with a flat-top filter shape can be realized, which is desirable for signal filtering in optical communications systems. On the other hand, when $t_2 = 0.742$, the CSLR resonator exhibits a flat-top group delay spectrum, which can be used as a Bessel filter for optical buffering.

**TABLE I. Definitions of waveguide and coupler parameters of the CSLR resonator**

| Waveguide | Length | Transmission factor[a] | Phase shift[b] |
|---|---|---|---|
| waveguide connecting $SLR_i$ to $SLR_{i+1}$ ($i = 1, 2, .., N-1$) | $L_i$ | $a_i$ | $\varphi_i$ |
| Sagnac loops in $SLR_i$ ($i = 1, 2, .., N$) | $L_{si}$ | $a_{si}$ | $\varphi_{si}$ |
| Coupler | Coupling length[c] | Field transmission coefficient[d] | Field cross-coupling coefficient[d] |
| couplers in $SLR_i$ ($i = 1, 2, .., N$) | $L_{ci}$ | $t_i$ | $\kappa_i$ |

[a] $a_i = exp(-\alpha L_i/2)$, $a_{si} = exp(-\alpha L_{si}/2)$, $\alpha$ is the power propagation loss factor.
[b] $\varphi_i = 2\pi n_g L_i/\lambda$, $\varphi_{si} = 2\pi n_g L_{si}/\lambda$, $n_g$ is the group index and $\lambda$ is the wavelength.
[c] $L_{ci}$ ($i = 1, 2, .., N$) are the straight coupling lengths shown in Fig.1. They are included in $L_i$.
[d] In our calculation, we assume $t_i^2 + \kappa_i^2 = 1$ for lossless coupling in all the directional couplers.

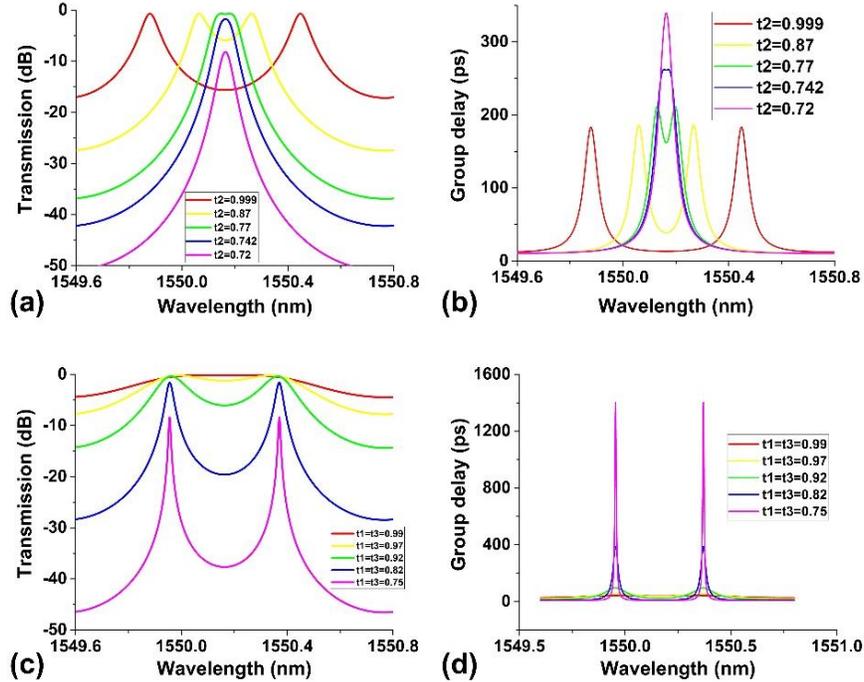

FIG. 2. (a) Calculated power transmission spectra of the CSLR resonator ($N = 3$) for various $t_2$ when $t_1 = t_3 = 0.87$. (b) Calculated group delay spectra of the CSLR resonator ($N = 3$) for various $t_2$ when $t_1 = t_3 = 0.87$. (c) Calculated power transmission spectra of the CSLR resonator ($N = 3$) for various $t_1 = t_3$ when $t_2 = 0.97$. (d) Calculated group delay spectra of the CSLR resonator ($N = 3$) for various $t_1 = t_3$ when $t_2 = 0.97$.

Figure 3(a) shows the calculated power transmission spectra of the CSLR resonators with different numbers of SLRs ($N$). It can be seen that as $N$ increases, the number of split resonances within one FSR also increases. For a CSLR resonator consisting of $N$ SLRs, the maximum number of split resonances within one FSR is $N-1$. In Fig. 3(b), we plot the calculated power transmission spectra of the CSLR resonator ($N = 8$) for different $t_1 = t_2 =...= t_8$. As $t_i$ ($i = 1, 2, ..., 8$) increases (i.e., the coupling strengths decrease), the bandwidth of the passband also increases, together with a decrease in insertion loss. In principle, the bandwidth of the passband is limited by the FSR of the CSLR resonator. The filter in Figs. 3(c) and (d) is designed for enhanced light trapping by introducing an additional $\pi/2$ phase shift along the centre FPC (i.e., $L_4$ for $N = 8$), which is similar to enhancing light trapping in photonic crystals by introducing defects [25]. With enhanced light trapping, there are increased time delays and enhanced light-matter interactions, which are useful in nonlinear optics and laser excitation [27]. In Fig. 3(c), one can see that there are central transmission peaks induced by an additional phase shift along $L_4$, which correspond to a group delay 2.1 times higher than that of the CSLR resonator without the additional phase shift in Fig. 3(d). This group delay can be increased further by using more cascaded SLRs. The filter in Fig. 3(e) is an 8th-order Butterworth filter with a flat-top filter shape. Figure 3(f) shows the designed optical filter with multiple transmission peaks in the spectrum. Each transmission peak has a high extinction ratio over 10 dB.

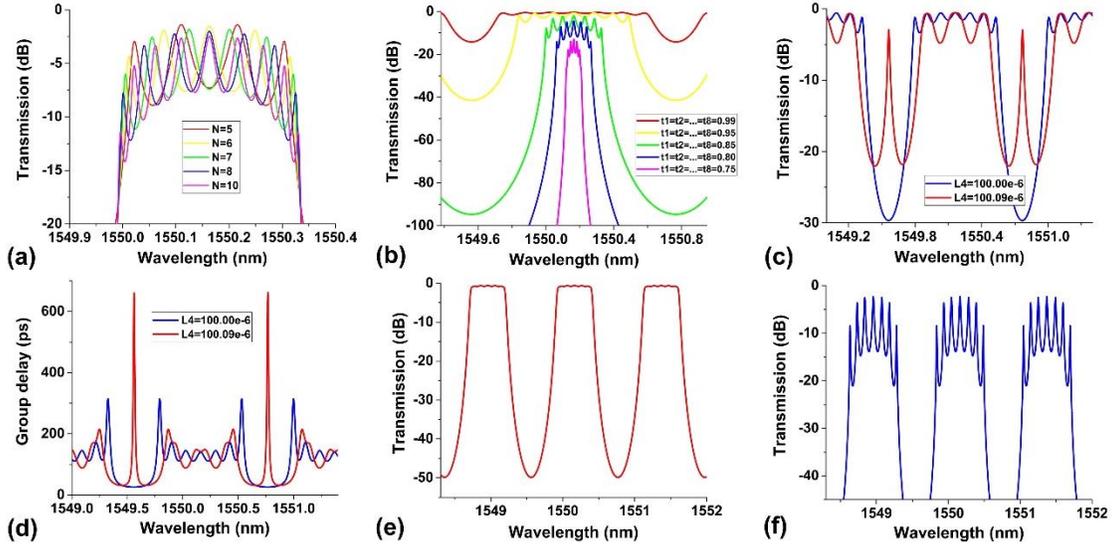

FIG. 3. (a) Calculated power transmission spectra of the CSLR resonator for various $N$ when $t_1 = t_2 =...= t_N = 0.85$. (c) Calculated power transmission spectra of the CSLR resonator ($N = 8$) for different $t_1 = t_2 =...= t_8$. (c) Calculated power transmission spectra of the CSLR resonator ($N = 8$) for enhanced light trapping. (d) Calculated group delay spectra of the CSLR resonator ($N = 8$) in (c). (e) Calculated power transmission spectra of 8th-order Butterworth filter based on the CSLR resonator ($N = 8$). (f) Calculated power transmission spectra of the CSLR resonator ($N = 8$) with multiple transmission peaks.

## 3. DEVICE FABRICATION AND CHARACTERIZATION

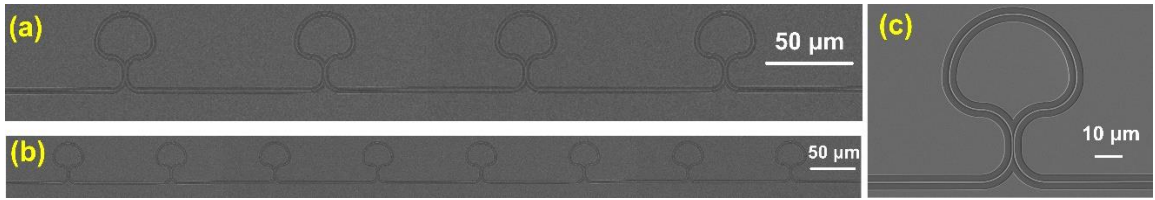

FIG. 4. (a) Micrograph for one of the fabricated CSLR resonators with four SLRs. (b) Micrograph for one of the fabricated CSLR resonators with eight SLRs. (c) Zoom-in micrograph for the SLR.

We fabricated a series of CSLR filters based on the above designs on an SOI wafer. The micrographs for the fabricated devices with four and eight SLRs are shown in Figs. 4(a) and (b), respectively. A zoom-in micrograph for the SLR is shown in Fig. 4(c). The normalized transmission spectra for the fabricated CSLR resonators with three SLRs ($N = 3$) are shown in Figs. 5(a) and (b) by the blue solid curves. The normalized spectra are then fit by the red dashed curves calculated based on the SMM. In Fig. 5(a), various mode splitting spectra of the fabricated devices with different $L_{c2}$ are obtained, which are consistent with the theory in Fig. 2(a). The measured spectra of the fabricated devices with different $L_{c1} = L_{c3}$ in Fig. 5(b) also agree well with the theory in Fig. 2(c).

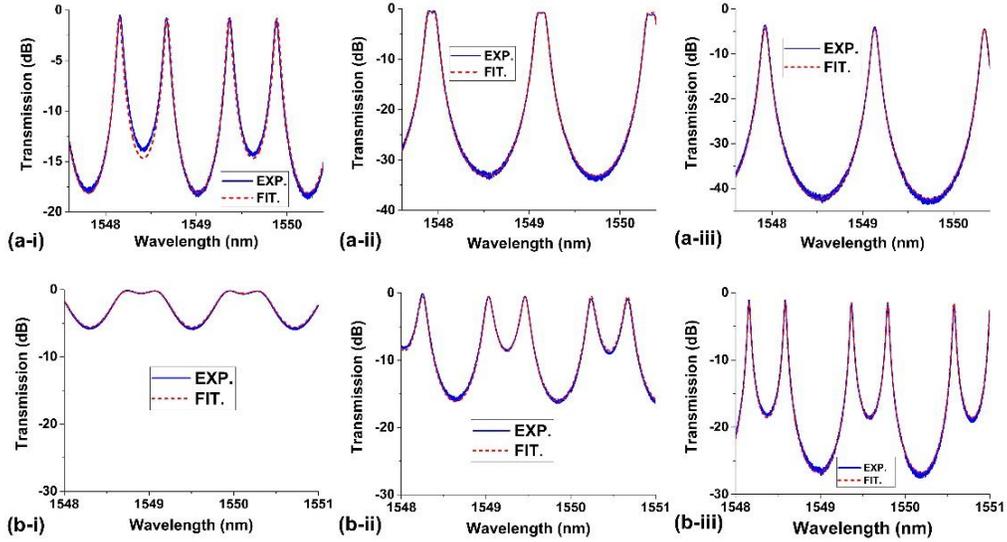

FIG. 5. (a) Measured (solid curve) and fit (dashed curve) transmission spectra of three fabricated CSLR resonators ($N = 3$) with different $L_{c2}$. (b) Measured (solid curve) and fit (dashed curve) transmission spectra of three fabricated CSLR resonators ($N = 3$) with different $L_{c1} = L_{c3}$.

The measured and fit transmission spectra of the fabricated CSLR resonators with four SLRs ($N = 4$) are shown in Fig. 6(a). Figures 6(b) − (d) show the measured and fit transmission spectra of the fabricated CSLR resonators with eight SLRs ($N = 8$). The device in Fig. 6(b) is designed for enhanced light trapping, and the measured transmission spectrum is similar to the calculated spectrum in Fig. 3(c). The measured filter shape in Fig. 6(b) exhibit a slight asymmetry, and this is because the additional phase shift along $L_4$ is not exactly $\pi/2$. By introducing thermo-optic micro-heaters or carrier-injection electrodes [26] along $L_4$ to tune the phase shift, the symmetry of the filter shape can be improved further. The device in Fig. 6(c) was designed to perform as an 8th-order Butterworth filter with a flat-top filter shape. As can be seen, the passband is almost flat, which is close to the calculated spectrum in Fig. 3(e). The slight unevenness of the top can be attributed to the discrepancies between the designed and practical coupling coefficients. Figure 6(d) shows the measured transmission spectrum with multiple resonant peaks. The minimum extinction ratio of the transmission peaks is ~7.8 dB, which is slightly lower than that in Fig. 3(f). This is mainly because the waveguide propagation loss of the fabricated devices ($\alpha$ = 64 m$^{-1}$) is slightly higher than we assumed in the calculation ($\alpha$ = 55 m$^{-1}$). By further reducing the propagation loss [33−36], as has been achieved in doped silica glass platforms such as Hydex, [37-85] higher extinction ratios of the split resonances can be obtained. These circuits with their high performance and low footprint will be highly useful for quantum optical integrated circuits [86-98] as well as for nonlinear optics involving 2D materials such as graphene oxide. [99-116]

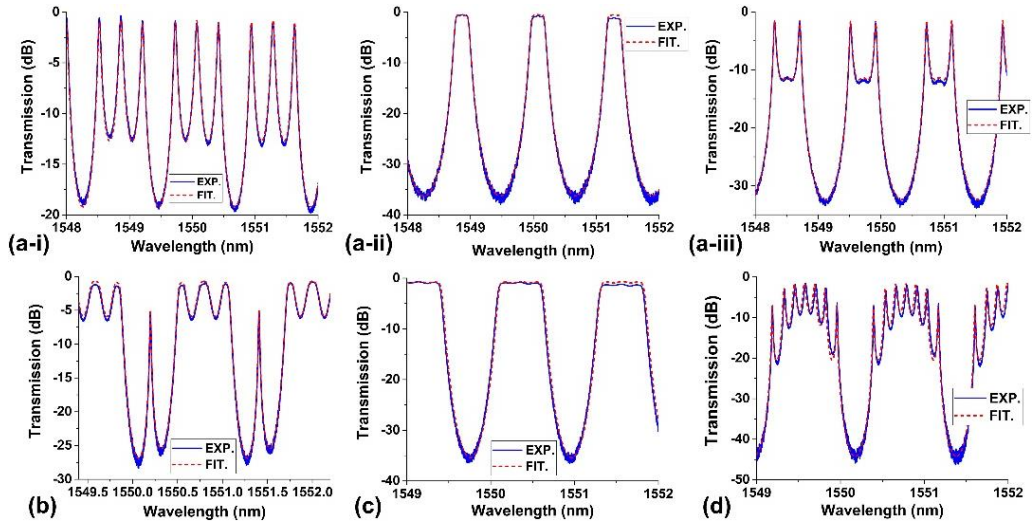

FIG. 6. (a) Measured (solid curve) and fit (dashed curve) transmission spectra of three fabricated CSLR resonators ($N = 4$) with different $L_{ci}$ ($i = 1, 2, 3, 4$). (b) Measured (solid curve) and fit (dashed curve) transmission spectra of a fabricated CSLR resonator ($N = 8$) for enhanced light trapping. (c) Measured (solid curve) and fit (dashed curve) transmission spectra of a fabricated CSLR resonator ($N = 8$) with flat-top filter shape. (d) Measured (solid curve) and fit (dashed curve) transmission spectra of a fabricated CSLR resonator ($N = 8$) with multiple split resonances.

## 4. CONCLUSION

In summary, we investigate mode splitting in standing-wave resonators formed by cascaded Sagnac loop reflectors (CSLRs) and demonstrate its use for engineering the spectral profile of integrated photonic filters. By changing the reflectivity of the Sagnac loop reflectors (SLRs) and the phase shifts along the connecting waveguides, we tailor the mode splitting to achieve a wide range of filter shapes for diverse applications including enhanced light trapping, flat-top filtering, Q factor enhancement, and signal reshaping. We present theoretical designs and experimentally demonstrate versatile filter shapes corresponding to diverse mode splitting conditions that agree well with theory.

**Conflict of Interest**

The authors declare that there are no conflicts of interest.